# Cognitive Mechanisms Underlying the Creative Process

**Liane M. Gabora**

## ABSTRACT

This paper proposes an explanation of the cognitive change that occurs as the creative process proceeds. During the initial, intuitive phase, each thought activates, and potentially retrieves information from, a large region containing many memory locations. Because of the distributed, content-addressable structure of memory, the diverse contents of these many locations merge to generate the next thought. Novel associations often result. As one focuses on an idea, the region searched and retrieved from narrows, such that the next thought is the product of fewer memory locations. This enables a shift from association-based to causation-based thinking, which facilitates the fine-tuning and manifestation of the creative work.

**Keywords**
Associative hierarchy, bisociation, brainstorm, concepts, conjunction, context, creativity, defocused attention, distributed representation, emergent features, evaluation, focus, generativity, idea, impossibilist creativity, intuition, variable focus.

## INTRODUCTION

What happens in the mind as a creative idea takes shape? This paper puts forth a theory of the cognitive mechanisms underlying creativity, elaborating on previous work [19, 20] to include the cognitive *change* that occurs as an idea transforms from inspiration to finished product. It assumes some form of homomorphism (rather than isomorphism) between the physical and the mental.

## STAGES OF THE CREATIVE PROCESS

The creative process has a long history of being divided into stages [3, 23, 60]. It is assumed that prior to the onset of a particular creative act, the creator has acquired the tools of the trade. The first stage is known as *preparation*, where the creator becomes obsessed with the problem, collects relevant data and traditional approaches to it, and perhaps attempts, unsuccessfully, to solve it.

During the second stage, termed *incubation*, the creator does not actively attempt to solve the problem, but unconsciously continues to work on it.

In the third stage, *illumination*, a possible surfaces to consciousness in a vague and unpolished form. Subjective and theoretical accounts of this phase of the creative process speak of discovering a previously unknown 'bisociation' [33], or underlying order. For example, Poincaré [50] claims that creative ideas "reveal unsuspected kinships between other facts well known but wrongly believed to be strangers to one another" (p. 115). The classic example is Kekule's discovery of the ring-shaped structure of the benzene molecule via a dream about a serpent biting its tail.

In the final stage, *verification*, the idea is worked into a form that can be proven and communicated to others.

Some argue that an incubation period may not be necessary [61], and that the creative process can be boiled down to a generative *brainstorming* stage followed by an evaluative *focusing* stage [9, 10]. Dennett suggests that the generative-evaluative process is cyclic; a new product is generated, evaluated, new goals are set, and the cycle is repeated.

## COGNITIVE MODES: ASSOCIATIVE AND ANALYTIC

The existence of two stages of the creative process is consistent with the widely-held view that there are two distinct forms of thought [9, 10, 30, 31, 46, 49, 54, 56]. The first is a suggestive, intuitive *associative mode* that reveals remote or subtle connections between items that are *correlated* but not necessarily *causally* related. This could yield a potential solution to a problem, though it may still be in a vague, unpolished form. The second form of thought is a focused, evaluative *analytic mode*, conducive to analyzing relationships of *cause and effect*. In this mode, one could work out the logistics of the solution and turn it into a form that is presentable to the world, and compatible with related knowledge or artifacts.

This suggests that creativity requires not just the capacity for both associative and analytic modes of thought, but also the ability to adjust the mode of thought to match the demands of the problem, and how far along one is in solving it. What cognitive mechanisms might underlie this?

## ARCHITECTURE OF THE MIND

Before we can piece together the cognitive mechanisms underlying creativity, we must briefly look at how episodes of experience, as well abstract items such as concepts, attitudes, and stories, are stored in memory.

### Memory is Sparse

The human mind would have to have more memory locations than the number of particles in the universe to store all the permutations of sound, colour, and so forth that the senses are capable of detecting. Thus, the number of memory locations is much smaller than the number of possible experiences.

This is illustrated schematically in Figure 1. Every vertex represents a possible constellation of stimulus properties that could be present in some experience we might have, and stored as an episode in memory. (A property might be something concrete such as 'blue' or more abstract such as 'honorable', or it may be something one would be unlikely to ever think of or come up with a word for.) Only a fraction of these constellations of properties is realized as *actual* memory locations (those with circles on them). The memory is therefore *sparse*. And in fact, only a fraction of *those* actually has some previous episode stored in them (those with black dots in the center).

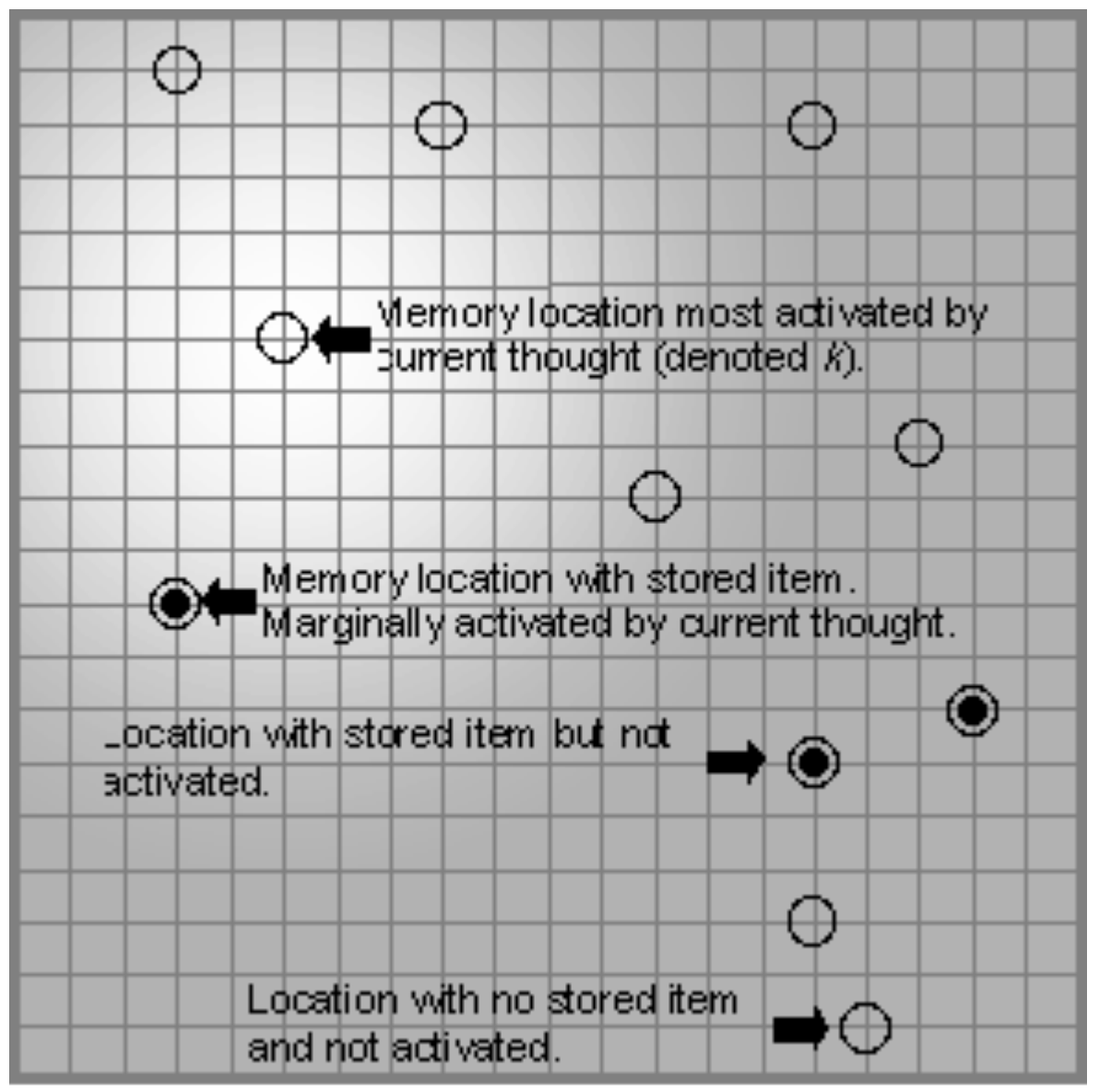

*Figure 1. Each vertex represents a possible memory location. Each black ring represents an actual location in a particular memory architecture or mind. The three rings with circles inside represent actual locations where an item has been stored. Degree of whiteness indicates degree of activation by current thought or experience. Activation is greatest for location k and falls with distance from k. In this case, only one location in the activated region has had something stored to it, and it is only marginally activated, so a reminding or retrieval event may or may not take place. If many memories had been stored in locations near k, they would blend to generate the next experience.*

## Memory is Distributed but Distributions are Constrained

If the mind stored each item in just one memory location as a computer does, then in order for an experience to evoke a reminding of a previous experience, it would have to be *identical* to that previous experience. And since the space of possible experiences is so vast that no two ever *are* exactly identical, this kind of organization would be somewhat useless.

In neural network models of cognitive processes, this problem can be solved by *distributing* the storage of an item across many memory locations [26, 32, 47, 62]. Likewise, each location participates in the storage of many items. However, in a *fully distributed* memory, where each item is stored in every memory location, the stored items interfere with one another. (This phenomenon goes by many names: 'crosstalk', 'superposition catastrophe', 'false memories', 'spurious memories' or 'ghosts' [17, 28, 29, 59] ).

The problem can be solved by *constraining* the distribution region. This is illustrated Figure 1; only a portion of the memory region (indicated by the degree of whiteness) gets activated. One way of constraining distributions in neural networks is to use a radial basis function, or RBF [24, 27, 36, 63]. Each input activates a hypersphere of memory locations, such that activation is maximal at the center *k* of the RBF and tapers off in all directions according to a (usually) Gaussian distribution of width s . Both *k* and s are determined in a training phase. The result is that one part of the network can be modified without affecting the capacity of other parts to store other patterns. A *spiky activation function* means that s is small. Therefore only those locations closest to *k* get activated, but they get activated a lot. A *flat activation function* means that s is large. Therefore locations relatively far from *k* still get activated, but no location gets *very* activated.

The mind is similarly constructed such that items in memory are distributed across assemblies of nerve cells [25, 37] but the distributions are constrained because assemblies are limited in size. Thus a given instant of experience activates not just *one* location in memory, nor does it activate *every* location to an equal degree, but activation is distributed across many memory locations, with degree of activation decreasing with distance from the most activated one, which we call *k*. The further a memory location is from *k*, the less activation it not only *receives* from the current stimulus experience but in turn *contributes* to the next instant of experience, and the more likely its contribution is cancelled out by that of other simultaneously activated locations. Following the neural network terminology, we say that the degree to which episodes and concepts are distributed is determined by the spikiness or flatness of the activation function. The flatter the activation function, the more distributed the memory.

The choice of *k* does not have to be determined in advance as it is in the neural network if memory locations differ in their capacity to detect and respond to different features. There is much evidence that this works through temporal coding [1, 6, 7, 8, 14, 35, 44, 45, 48, 53, 58, for reviews see 11, 51]. As Cariani points out, temporal coding drastically simplifies the problem of how the brain coordinates, binds, and integrates information. Different features or stimulus dimensions are carried by different frequencies like a radio broadcast system, and each memory location is attuned to respond to a slightly different frequency, or set of frequencies. Thus, *k* arises naturally due to the differential effect of the stimulus on the various memory locations. The greater the number of stimulus frequencies impacting the memory architecture, the greater the number of memory locations that respond.

## Memory is Content Addressable

Content addressability refers to the fact that there is a systematic relationship between the content of an experience (not just as the subject matter, but the qualitative feel of it) and the memory locations where it gets stored (and from which material for the next instant of experience is evoked). In a computer, this one-to-one correspondence arises naturally because each possible input a unique address in memory. Retrieval is thus simply a matter of looking at the address in the address register and fetching the item at the specified location. The distributed nature of human memory prohibits this, but content addressability is still achievable as follows. Memory locations are contained in nerve cells called neurons. A given experience induces a chain reaction such that some neurons are inhibited and others excited. For an experience to be engraved to a certain memory location, a particular pattern of activation must occur. The 'address' of a memory location is thus the pattern(s) of activation that lead it to be affected.

## Associative Richness

We have looked at three aspects of the architecture of memory: it is sparse, distributed (yet distributions are constrained), and content-addressable. In such a memory architecture, if the regions where two stored episodes or abstract concepts overlap, it means they share one or more common features or properties. The relationship between memories and concepts stored in overlapping regions of memory is therefore one of *correlation*rather than *causation* . Now let us look at why this turns out to be important for creativity.

Martindale [39] has identified a cluster of psychological attributes associated with creativity which includes defocused attention [12, 13, 43], and high sensitivity [39, 40], including sensitivity to subliminal impressions; that is, stimuli that are perceived but of which we are not *conscious* of having perceived [57].

Another characteristic of creative individuals is that they have *flat associative hierarchies* [42]. The steepness of an individual's associative hierarchy is measured experimentally by comparing the number of words that individual generates in response to stimulus words on a word association test. Those who generate only a few words in response to the stimulus have *steep* associative hierarchies. Those who generate many have *flat*associative hierarchies. One can also refer to this as *associative richness* . Thus, once such an individual has run out of the more usual associations (*e.g* . <u>chair</u> in response to <u>table</u>), unusual ones ( *e.g*. <u>elbow</u> in response to<u>table</u>) come to mind.

The experimental evidence that creativity is associated with not just flat associative hierarchies but also defocused attention and heightened sensitivity suggests that associative richness stems from a tendency to perceive more of the detail of a stimulus or situation. One includes in ones' internal representation of the stimulus situation features that are less central to the concept that best categorizes it, features that may in fact make it defy straightforward classification as strictly an instance of one concept or another. This could be accomplished through a tendency toward a flat activation function. Experiences get more widely etched into memory, thus the storage regions for episodes and concept overlap more, resulting in greater potential for associations to be found amongst them.

## A Stream of Thought

Since content addressability ensures that items with related meanings get stored in overlapping locations, one naturally retrieves items that are similar or relevant to the current experience. As a result, the entire memory does not have to be searched in order for, for example, one person to remind you of another. It is because the size of the region of activated memory locations must falls midway between the two possible extremes--not distributed and fully distributed--that one can generate a stream of coherent yet potentially creative thought. The current thought or experience activates a certain region of memory. Episodes or concepts stored in the locations in this region provide the 'ingredients' for the next thought. This next thought is slightly different, so it activates and retrieves from a slightly different region, and so forth, recursively.

In a state of defocused attention or heightened sensitivity to detail, stimulus properties that are less directly relevant to the current goal get encoded in memory. Since more features of attended stimuli participate in the process of storing to and evoking from memory, more memory locations are activated and participate in the encoding of an instant of experience and release of 'ingredients' for the next instant. The more memory locations activated, the more they in turn activate, and so on; thus streams of thought tend to last longer. So if a stimulus does manage to attract attention, it will tend to be more thoroughly assimilated into the matrix of associations that constitutes the worldview, and more time is taken to settle into any particular interpretation of it. We refer to this as a state of *conceptual fluidity*. In such a state, new stimuli are less able to compete with what has been set in motion by previous stimuli, *i.e.* the memory network plays a larger role in conscious experience.

Another interesting consequence of a flat activation function is that an episode or concept that lies relatively far from the one that best captures the properties of the current thought, but that at least lies within the hypersphere of activated memory locations, can 'pull' the next thought quite far from the one that preceded and evoked it. Thus there is an increased probability that one thought will lead in a short period of time to a seemingly unrelated thought; consecutive instants are less correlated. So the worldview is penetrated more deeply, but also traversed more quickly.

## VARIABLE FOCUS AS THE KEY TO CREATIVITY

There is in fact a considerable body of research suggesting that creativity is associated with, not just high conceptual fluidity, nor just extraordinary control, but both [2, 15, 16, 18, 52, 55]. As Feist [16] puts it: "It is not unbridled psychoticism that is most strongly associated with creativity, but psychoticism tempered by high ego strength or ego control. Paradoxically, creative people appear to be simultaneously very labile and mutable and yet can be rather controlled and stable" (p. 288). He notes that, as Barron [2] said over 30 years ago: "The creative genius may be at once naïve and knowledgeable, being at home equally to primitive symbolism and rigorous logic. He is both more primitive and more cultured, more destructive and more constructive, occasionally crazier yet adamantly saner than the average person" (p. 224). There is also evidence of an association between creativity and high variability in physiological measures of arousal such as heart rate [4], spontaneous galvanic skin response [38], and EEG alpha amplitude [39, 41].

Knowing that creativity is associated with *both* conceptual fluidity on the one hand, and focus or control on the other, puts us in a good position to posit an underlying mechanism: the capacity to spontaneously adjust the spikiness of the activation function in response to the situation. Each new instant of thought can touches more or fewer memory locations, depending on the nature of the problem, and how far along one is in the process of solving it. We can refer to this as the capacity for *variable focus*. Let us now look in more detail at how it could explain what happens during the creative process from initial brainstorming stage to fine-tuning of the finished product.

## Brainstorming and insight

Let us now consider what happens in the mind of an artist or scientist who does something particularly creative. The first response to a problem, perceived inconsistency, or desire to express oneself or generate something asthetically pleasing, may be a rational or deductive approach. When this doesn't work, it seems likely that there is a tendency to brainstorm; temporarily 'loosen' one's internal model of reality, weaken inter-concept relationships so as to allow new insights to more readily percolate through and exert the needed revolutionary impact. One becomes receptive to new ways of perceiving the world. How might this happen?

Dennett [10] believes that the generative component of the creative process operates randomly. Campbell [5] claims it operates through a process of blind variation. He specifically states that by 'blind' he does not mean random, and is emphatic that it is not *causal*, though how it does work he does not say. Nevertheless, the notion that the generative stage is not random yet not causal is readily explainable given the attributes of memory described above. In a state of defocused attention and heightened sensitivity, more features of the stimulus situation or concept under consideration get processed. Thus the more memory locations the current instant of experience gets stored to; the activation function is flat, as in Figure 2.

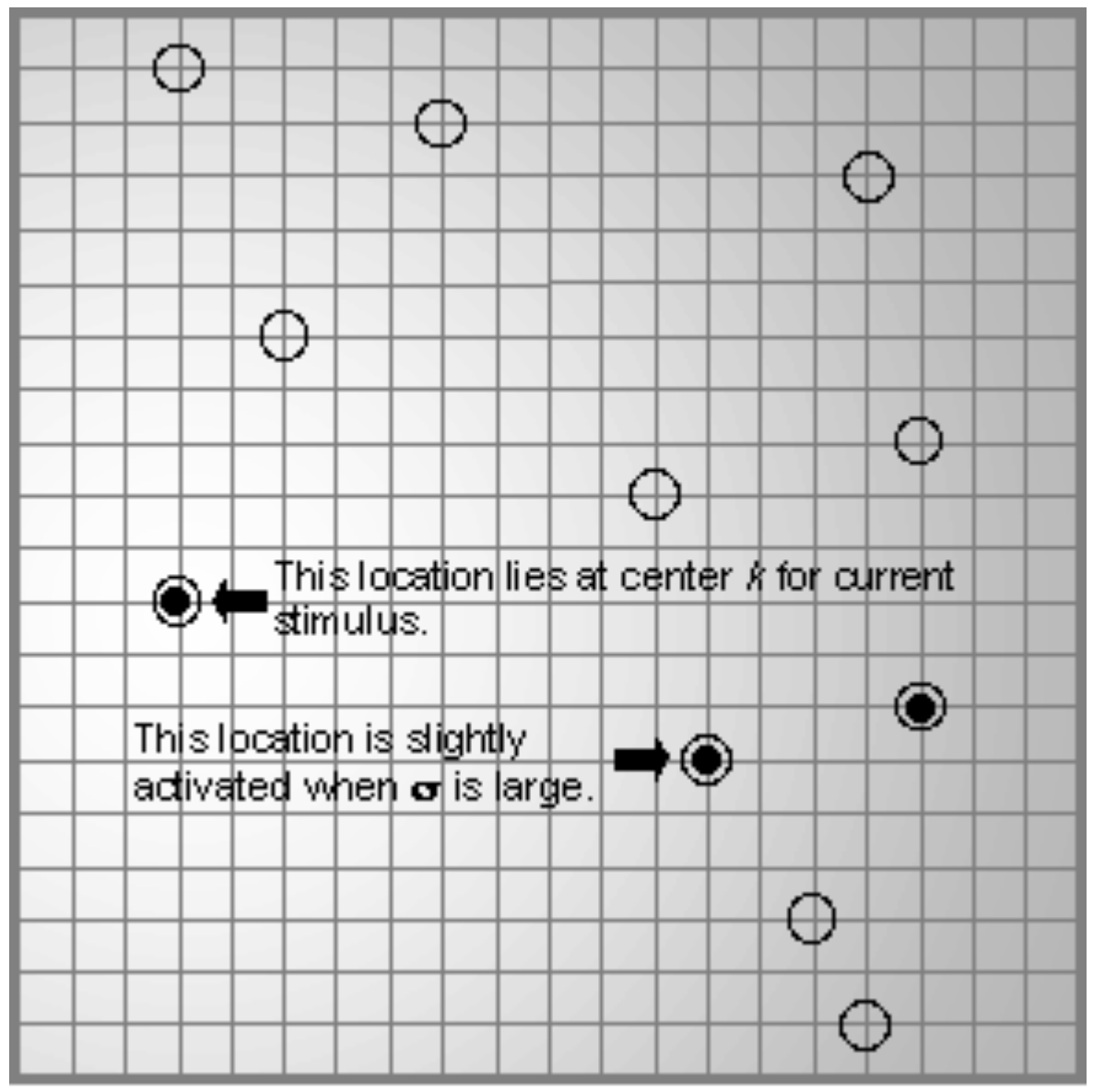

*Figure 2. Schematic representation of region of memory activated and retrieved from during the associative stage of creative process. Stored items in all the memory locations in whitened region will blend to generate the next instant of thought.*

The flat activation function results in a greater likelihood of 'catching' a stored episode or concept that isn't usually associated with the experience that evoked it. The new idea arises in an unpolished form; it exists in a state of *potentiality*, in the sense that the newly identified relationship could be resolved different ways depending on the contexts one encounters, both immediately, and down the road. For example, consider the cognitive state of the person who thought up the idea of building a snowman. It seems reasonable that this involved thinking of <u>snow</u> not just in terms of its most typical features such as `cold' and `white', but also the less typical feature `moldable'. At the instant of inventing snowman there were many ways of resolving how to give it a nose. However, perhaps because the inventor happened to have a carrot handy, the concept snowman has come to acquire the feature `carrot nose'.

Thus new ideas arise through a sort of 'conceptual meltdown', in that the details of episodes or the meanings of concepts merge or blend into one another more than usual, such that they are more readily recombined to give something unique.

## Focusing the Creative Idea

In the short run, a wide activation function is conducive to creativity because it provides a high probability of 'catching' new combinations of properties by reconstructing unusual blends of stored items. But maintaining it indefinitely would be untenable since the relationship between one thought and the next can be so remote that a stream of thought lacks continuity. Thus, once the overall framework of a unique idea has been painted in the broad strokes, one goes from a state of mind that is more likely to simultaneously evoke items that are correlated and therefore stored in overlapping memory locations, to a state of mind that is more conducive to establishing relationships of causation. This may happen by gradually narrowing the region of memory that gets activated, such that fewer memory locations are activated, as in Figure 3. Fewer locations release their contents to 'participate in' the formation of the next though, thus affording one finer control over what concepts gets evoked. Thought becomes focused and logical; access to remote associations, and the ensuing generation of strange new combinations, would at this point be a distraction.

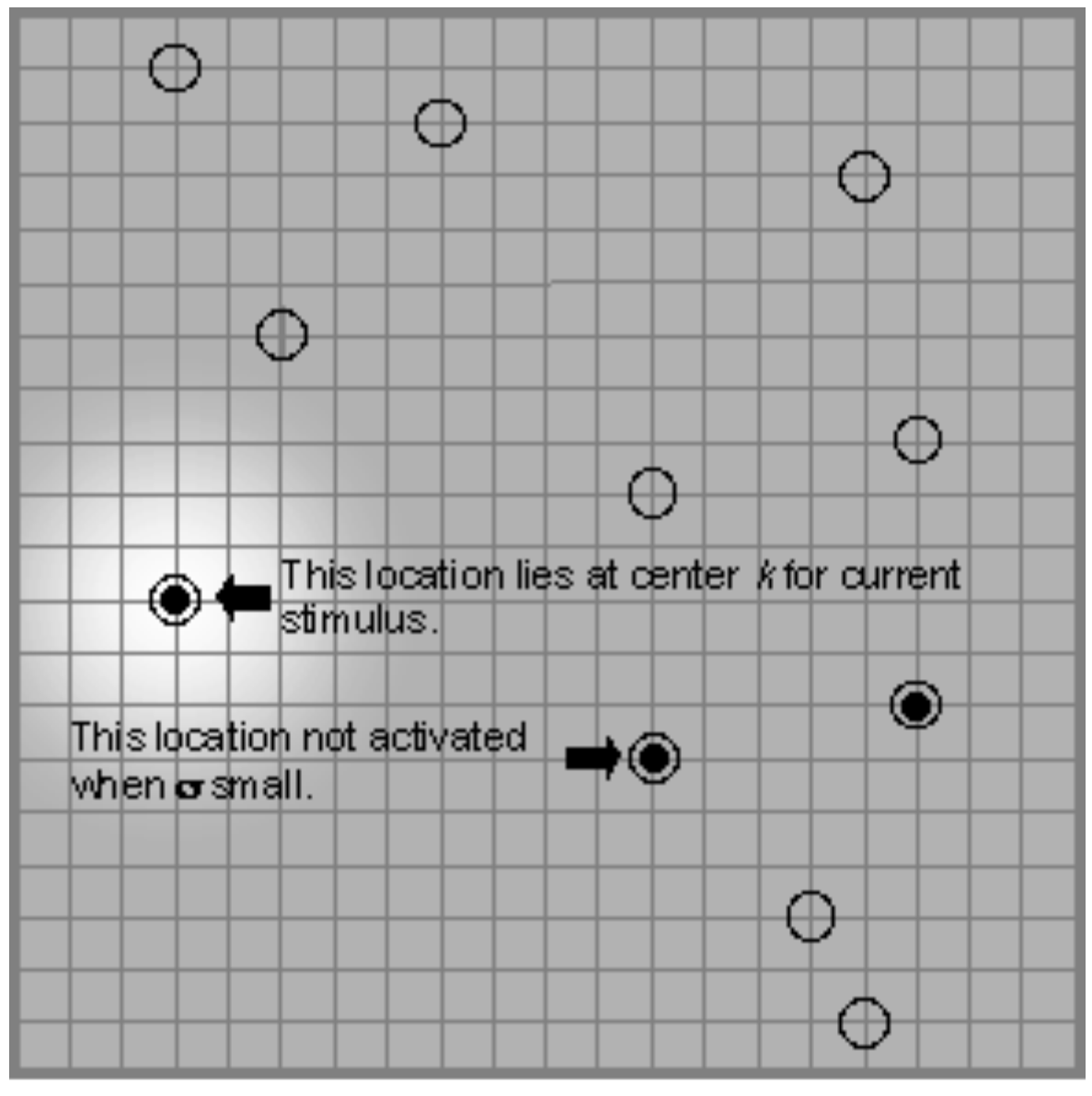

*Figure 3. Schematic representation of region of memory activated and retrieved from during the more analytic stage of the creative process. The activation function is spiky.*

Through such focusing, one slowly settles on a 'draft' of the idea that incorporates aspects that are relevant, pleasing, or useful, while weeding out aspects that are irrelevant, distasteful, or misleading. In the process, the idea becomes grounded more firmly in consensus reality, so that when it is born it is more widely understandable and less vulnerable to attack.

## MODELING THE CREATIVE PROCESS

Is it possible to mathematically model the creative process? One big stumbling block is that a creative idea often possesses features which are said to be *emergent:* not true of the constituent ideas of which it was composed. For example, the concept snowman has as a feature or property 'carrot nose', though neither snow nor man does). Most mathematical formalisms are not only incapable of *predicting* what sorts of features will emerge (or disappear) in the conjunctive concept, but they do not even provide a place in the formalism for the gain (or loss) of features. This problem stems back to a limitation of the mathematics underlying not only representational theories of concepts (as well as compositional theories of language) but all classical physical theories. The mathematics of classical physics only allows one to describe a composite or joint entity by means of the product state space of the state spaces of the two subentities. Thus if $X_1$ is the state space of the first subentity, and $X_2$ the state space of the second, the state space of the joint entity is the Cartesian product space $X_1 \times X_2$. For this reason, classical physical theories cannot describe the situation wherein two entities generate a new entity with properties not strictly inherited from its constituents.

One could try to solve the problem *ad hoc* by starting all over again with a new state space each time there appears a state that was not possible given the previous state space; for instance, every time a conjunction comes into existence. However, this happens every time one generates a sentence that has not been used before, or even uses the same sentence in a slightly different context. Another possibility would be to make the state space infinitely large to begin with. However, since we hold only a small number of items in mind at any one time, this is not a viable solution to the problem of describing what happens in cognition. This problem is hinted at by Boden [3], who uses the *term impossibilist creativity* to refer to creative acts that not only *explore* the existing state space but *transform* that state space; in other words, it involves the spontaneous generation of new states with new properties.

However, this sort of problem is addressed by mathematical formalisms originally developed for quantum mechanics, as follows. When quantum entities combine, they do not stay separate as classical physical entities tend to do, but enter a state of *entanglement*. If $H_1$ is the Hilbert space describing first subentity and $H_2$ the Hilbert space describing second, the state space of the joint entity is described by the tensor product of the two Hilbert spaces. The tensor product always allows for the emergence of new states--specifically the entangled states--with new properties.

Because of the linearity of Hilbert space, the mathematics of pure quantum mechanics is too limited to begin to describe how concepts combine in the mind to form new ideas. However, this (and other) limitations are overcome using a generalization of the pure quantum formalism known as the *state context property system*, or SCOP formalism. This formalism is being used to articulate a theory of creativity by treating complex ideas as conjunctions as *concepts in the context of one another* [21, 22]. Whereas the role of context is generally neglected, as we see it, ideas *require* a context to actualize them in a thought or experience. The SCOP formalism enables us to explicitly incorporate the context that elicits a reminding of a concept, and the change of state this induces in the concept (possibly transforming it into something new and creative) into the formal description of the concept itself. A concept is viewed not as a fixed mental representation, but as a source of potentiality which predisposes it to dynamically attract context-specific cognitive states (both concrete stimulus experiences and imagined or counterfactual situations) into a certain subspace of conceptual space. Interaction with the context (the stimulus or situation) causes a concept to `collapse' to an instantiated form of it. Thus a concept cannot be described in a context-independent manner (except as a superposition of every possible context-driven instantiation of it). In this view, not only does a concept give meaning to a stimulus or situation, but the situation evokes meaning in the concept, and when more than one is active they evoke meaning in each other. Each of the two concepts in a conjunction constitutes a context for the other that `slices through' it at a particular angle, thereby mutually actualizing one another's potentiality in a specific way. The stimulus situation plays the role of the measurement in physics, acting as context that induces a change of the cognitive state from superposition state to collapsed state. The collapsed state is more likely to consist of a conjunction of concepts for associative than analytic thought because more stimulus or concept properties take part in the collapse. As a metaphorical explanatory aid, if concepts were apples, and the stimulus a knife, then the qualities of the knife would determine not just which apple to slice, but which direction to slice through it. Changing the knife (the context) would expose a different face of the apple (elicit a different version of the concept). And if the knife were to slash through several apples (concepts) at once, we might end up with a new kind of apple (a conjunction).

## SUMMARY

We have looked at a cognitive mechanism to explain what happens in the mind during the course of the creative process. Until a creative insight has been obtained, one is in an intuitive, brainstorming state of mind, and the activation function is wide. The creative insight may take many forms: for example, an invention or scientific theory, story or myth, a way or moving, acting, or accomplishing something, or a way of portraying relationship and emotion artistically. At this point the new idea is still vague and needs to come into focus. One reflects on an idea by reflecting it back into the memory with an increasingly spiky activation function, seeing what it evokes back into awareness, and repeating the process until it comes into focus. By taking what is retrieved from memory (which may consist of many items blended together) and feeding the most promising properties of this construction back at the memory, seeing what is then retrieved, and so forth, certain properties get abstracted. The initially unfocused idea eventually turns into one that can solve the problem at hand, account for the inconsistency, or convey the desired relations and emotions.

The mathematical modeling of the creative process is a difficult endeavor, partly because the new idea often has properties that were not present in the constituent ideas or concepts that went into the making of it. It is possible to use a mathematical formalism that was originally devised partly to cope with the problems of context and emergent properties in the quantum world. In this model, interaction with a context (the stimulus or situation) causes a concept to `collapse' to a possibly new instantiated form of it. The collapsed state is more likely to consist of a conjunction of concepts for associative than analytic thought because, due to the flat activation function, more stimulus or concept properties take part in the collapse.

## ACKNOWLEDGMENTS

I would like to acknowledge the support of Grant FWOAL230 of the Flemish Fund for Scientific Research. The part at the end about modeling the contextual aspects of cognition using formalisms originally developed for description of contextuality in physics was developed with Diederik Aerts, and is more fully described in our joint papers.

## REFERENCES

Abeles, M. & Bergman, H. Spatiotemporal firing patterns in the frontal cortex of behaving monkeys. *Journal of Neurophysiology* 70 **,** 4 (1993), 1629-1638.

Barron, F. *Creativity and Psychological Health*, Van Nostrand, 1963.

Boden, M. *The Creative Mind: Myths and Mechanisms*. Weidenfeld & Nicolson. Revised edition, Cardinal, 1990/1992.

Bowers, K.S. & Keeling, K.R. Heart-rate variability in creative functioning. *PsychologicalReports* 29 (1971), 160-162.

Campbell, D. Evolutionary Epistomology. In *Evolutionary Epistomology, Rationality, and the Sociology of Knowledge*, eds. G. Radnitzky &. W.W. Bartley III, Open Court, LaSalle IL, 1987.

Campbell, F.W. & Robson, J.G. Application of Fourier analysis to the visibility of gratings. *Journal of Physiology*, 197 (1968), 551-566.

Cariani, P. As if time really mattered: temporal strategies for neural coding of sensory information. In *Origins: Brain and self-organization* , ed. K. Pribram. Erlbaum, Hillsdale NJ, 1995, 161-229.

Cariani, P. Temporal coding of sensory information. In *Computational neuroscience: Trends in research 1997*, ed. J.M. Bower. Plenum, Dordrecht, Netherlands, 1997, 591-598.

Dartnell, T. Artificial intelligence and creativity: An introduction. *Artificial Intelligence and the Simulation of Intelligence Quarterly* 85 (1993).

Dennett, D. *Brainstorms: Philosophical Essays on Mind and Psychology* , Harvester Press, 1978.

De Valois, R.L. & De Valois, K.K. *Spatial Vision.* Oxford University Press, Oxford UK, 1988.

Dewing, K. & Battye, G. Attentional deployment and non-verbal fluency. *Journal of Personality and Social Psychology* 17 (1971), 214-218.

Dykes, M. & McGhie, A. A comparative study of attentional strategies in schizophrenics and highly creative normal subjects. *British Journal of Psychiatry* 128 (1976), 50-56.

Emmers, R. *Pain: A Spike-Interval Coded Message in the Brain.* Raven Press, Philadelphia PA, 1981.

Eysenck, H.J. *Genius: The Natural History of Creativity*, Cambridge University Press, Cambridge UK, 1995.

Feist, G.J. The influence of personality on artistic and scientific creativity. In: *Handbook of Creativity,* ed. R. J. Sternberg, Cambridge University Press, Cambridge UK (1999), 273-296.

Feldman, J.A. & Ballard, D.H. Connectionist models and their properties. *Cognitive Science* 6 (1982), 205-254.

Fodor, E.M. Subclinical manifestations of psychosis-proneness, ego-strength, and creativity. *Personality and Individual Differences* 18 (1995), 635-642.

Gabora, L. The beer can theory of creativity. In *Creative Evolutionary Systems,* eds. P. Bentley & D. Corne, Morgan Kauffman (2002), 147-161. Available athttp://cogprints.soton.ac.uk/documents/disk0/00/00/09/76/

Gabora, L. Toward a theory of creative inklings. In (R. Ascott, Ed.) *Art, Technology, and Consciousness*, Intellect Press (2000), 159-164. Available at http://cogprints.soton.ac.uk/documents/disk0/00/00/08/56/

Gabora, L. & Aerts, D. Contextualizing concepts. *Proceedings of the 15th International FLAIRS Conference* (Pensacola Florida, May 2002), American Association for Artificial Intelligence, 148-152.

Gabora, L. & Aerts, D. Contextualizing concepts using a mathematical generalization of the quantum formalism. *Journal of Experimental and Theoretical Artificial Intelliegence*, to appear in special issue on concepts and categories (2002). http://www.vub.ac.be/CLEA/liane/papers/jetai.pdf

Hadamard, J. *The Psychology of Invention in the Mathematical Field* . Princeton University Press, Princeton NJ, 1949.

Hancock, P.J.B., Smith, L.S. & Phillips, W.A. A biologically supported error-correcting learning rule. *Neural Computation* 3, 2 (1991), 201-212.

Hebb, D.O. *The Organization of Behavior*. Wiley, 1949.

Hinton, G., McClelland, J.L. & Rummelhart, D.E. Distributed representations. In *Parallel distributed processing: Explorations in the microstructure of cognition*, eds. Rummelhart, D. E. & J. L. McClelland. MIT Press, Cambridge MA, 1986.

Holden, S.B. & Niranjan, M. Average-case learning curves for radial basis function networks. *Neural Computation* 9, 2 (1997), 441-460.

Hopfield, J.J. Neural networks and physical systems with emergent collective computational abilities', *Proceedings of the National Academy of Sciences* 79 (1982), 2554-2558.

Hopfield, J.J., Feinstein, D.L. & Palmer, R.G. "Unlearning" has a stabilizing effect in collective memories. *Nature* 304 (1983), 158-159.

James, W. *The Principles of Psychology*. Dover, New York, 1890/1950.

Johnson-Laird, P.N. *Mental Models*. Harvard University Press, Cambridge MA, 1983.

Kanerva, P. *Sparse Distributed Memory*, MIT Press, Cambridge MA, 1988.

Koestler, A. *The Act of Creation*. Macmillan, 1964.

Lestienne, R. Determination of the precision of spike timing in the visual cortex of anesthetized cats. *Biology and Cybernetics* 74 (1996), 55-61.

Lestienne, R. amd Strehler, B.L. Time structure and stimulus dependence of precise replicating patterns present in monkey cortical neuron spike trains. *Neuroscience* (April 1987).

Lu, Y.W., Sundararajan, N. & Saratchandran, P. A sequential learning scheme for function approximation using minimal radial basis function neural networks. *Neural Computation* 9, 2 (1997) 461-478.

Marr, D. A theory of the cerebellar cortex. *Journal of Physiology* , 202 (1969), 437-470.

Martindale, C. Creativity, consciousness, and cortical arousal. *Journal of Altered States of Consciousness* 3 (1977), 69-87.

Martindale, C. Biological bases of creativity. In *Handbook of Creativity* , ed. R. J. Sternberg, Cambridge University Press, Cambridge UK (1999), 137-152.

Martindale, C. & Armstrong, J. The relationship of creativity to cortical activation and its operant control. *Journal of Genetic Psychology* 124 (1974), 311-320.

Martindale, C. & Hasenfus, N. EEG differences as a function of creativity, stage of the creative process, and effort to be original. *Biological Psychology* 6 (1978), 157-167.

Mednick, S.A. The associative basis of the creative process. *Psychological Review* 69 (1962), 220-232.

Mendelsohn, G.A. Associative and attentional processes in creative performance. *Journal of Personality* 44 (1976), 341-369.

Metzinger, T. Faster than thought: Holism, homogeneity, and temporal coding. In *Conscious Experience*, ed. T. Metzinger. Schoningh/Academic Imprint, Thorverton U.K., 1995.

Mountcastle, V. Temporal order determinants in a somatosthetic frequency discrimination: Sequential order coding. *Annals of the New York Academy of Science* 682 (1993), 151-170.

Neisser, U. The multiplicity of thought. *British Journal of Psychology* 54 (1963), 1-14.

Palm, G. On associative memory. *Biological Cybernetics* 36 (1980), 19-31.

Perkell, D.H. & Bullock, T.H. Neural coding. *Neurosciences Research Program Bulletin* 6, 3 (1968), 221-348.

Piaget, J. *The Language and Thought of the Child*. Routledge & Kegan Paul, London, 1926.

Poincare, H. *The Foundations of Science.* Science Press, Lancaster PA, 1913.

Pribram, K.H. *Brain and Perception: Holonomy and Structure in Figureal Processing.* Erlbaum, Hillsdale NJ, 1991.

Richards, R.L., Kinney, D.K., Lunde, I., Benet, M., & Merzel, A. Creativity in manic depressives, cyclothymes, their normal relatives, and control subjects. *Journal of Abnormal Psychology* 97 (1988), 281-289.

Riecke, F. & Warland, D. *Spikes: Exploring the Neural Code* . MIT Press, Cambridge, MA, 1997.

Rips, L.J. Necessity and natural categories. *Psychological Bulletin* 127, 6 (2001) 827-852.

Russ, S.W. *Affect and Creativity*. Erlbaum, Hillsdale NJ, 1993.

Sloman, S. The empirical case for two systems of Reasoning. *Psychological Bulletin* 9, 1 (1996), 3-22.

Smith, G.J.W. & Van de Meer, G. Creativity through psychosomatics. *Creativity Research Journal* 7 (1994), 159-170.

Stumpf, C. Drug action on the electrical activity of the hippocampus. *International Review of Neurobiology* 8 (1965), 77-138.

Von der Malsburg, C. Am I thinking assemblies? In *Proceedings of the 1984 Trieste Meeting on Brain Theory*, ed. G. Palm & A. Aertsen. Springer-Verlag, Berlin, 1986.

Wallas, G. *The Art of Thought*. Harcourt, Brace & World, New York, 1926.

Weisberg, R. *Creativity: Genius and Other Myths*. Freeman Press, New York, 1986.

Willshaw, D.J. Holography, associative memory, and inductive generalization. In *Parallel Models of Associative Memory*, eds. G.E. Hinton & J.A. Anderson, Lawrence Earlbaum Associates (1981), 83-104.

Willshaw, D. J. & Dayan, P. Optimal plasticity from matrix memory: What goes up must come down. *Journal of Neural Computation* 2 (1990), 85-93.